\documentclass[a4paper]{jpconf}

\usepackage{graphicx}

\begin{document}

\title
  {Measurement of direct photons in $\sqrt{s_{NN}} = 200$\,GeV $p+p$ and
  Au+Au collisions with the PHENIX Experiment at RHIC}

\author{Stefan Bathe for the PHENIX Collaboration}

\address{University of California, Riverside, CA, USA}

\ead{bathe@bnl.gov}

\begin{abstract}
The measurement of direct photons in $\sqrt{s_{NN}} = 200$\,GeV $p+p$
and Au+Au collisions is presented.  The signal is compared to NLO pQCD
calculations, which, in case of Au+Au, are scaled with the number of
underlying nucleon-nucleon collisions.  The agreement of the
calculation with the data in both cases confirms the scaling of hard
processes with the number of nucleon-nucleon collisions and supports
the explanation of the earlier-observed pion suppression as a
final-state effect.

\end{abstract}

\section{Introduction}

Direct photons are a unique probe of the hot and dense matter created
at RHIC: they allow access to the initial, thermalized state of the
nuclear collision.  Their measurement, however, is challenging.  One
has to cope with a large background from hadronic decays.

While several definitions of direct photons are used, we pragmatically
attribute this term to all photons that are not decay photons.  The
main direct photon production processes are quark-antiquark
annihilation and quark-gluon Compton scattering
\cite{Peitzmann:2001mz}.  In addition, direct photons are produced
through fragmentation of hard partons. These are also called
bremsstrahlung photons.

Direct photon measurements in $p+p$ collisions provide a superb test
of pQCD.  They probe the gluon distribution function.  At RHIC they
can also probe the spin gluon distribution function
\cite{Saito:1998cx} in the polarized-proton program.  In contrast to
hadron measurements, their interpretation does not suffer much from
uncertainties on the fragmentation function.  Furthermore, they
provide a constraint on the hard-scattering contribution to direct
photon production in Au+Au collisions.

In heavy ion collisions, thermal direct photons allow, in principle,
measurement of the temperature of the collision system in its hottest
phase.  Also, hard direct photons serve as a crucial baseline for the
interpretation of the earlier observed high-$p_T$ hadron suppression
\cite{Adler:2003qi}.  Interactions of hard partons with the medium
provide an additional source of direct photons, either through
annihilation and Compton scattering \cite{Fries:2002kt} or through
medium-induced bremsstrahlung \cite{Jeon:2002mf}.

With its high-resolution, highly-segmented electromagnetic calorimeter
PHENIX \cite{Adcox:2003zm} has an excellent capability to measure
photons.  PHENIX has made precision measurements of neutral pions and
$\eta$ mesons up to transverse momenta of 15 GeV/c.  With this crucial
measurement of the direct photon background, PHENIX has extracted
direct photons in $p+p$ and Au+Au collisions at $\sqrt{s_{NN}} =
200$\,GeV.

\section{Direct Photon Analysis}

\begin{figure}[t]
\begin{minipage}{18pc}
\begin{center}
\includegraphics[width=16pc]{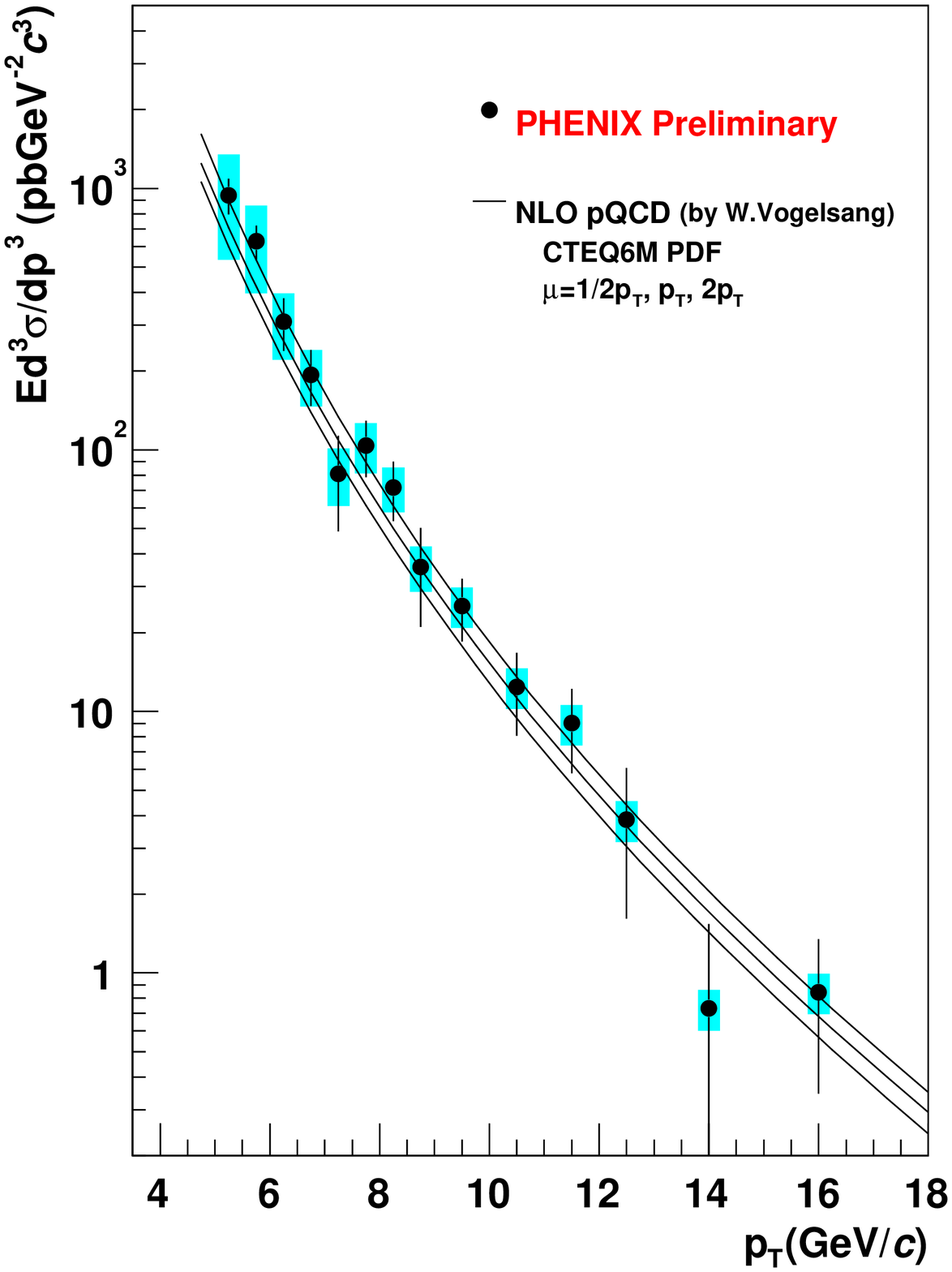}
\end{center}
\end{minipage}\hspace{2pc}%
\begin{minipage}{18pc}
\begin{center}
\includegraphics[width=16pc]{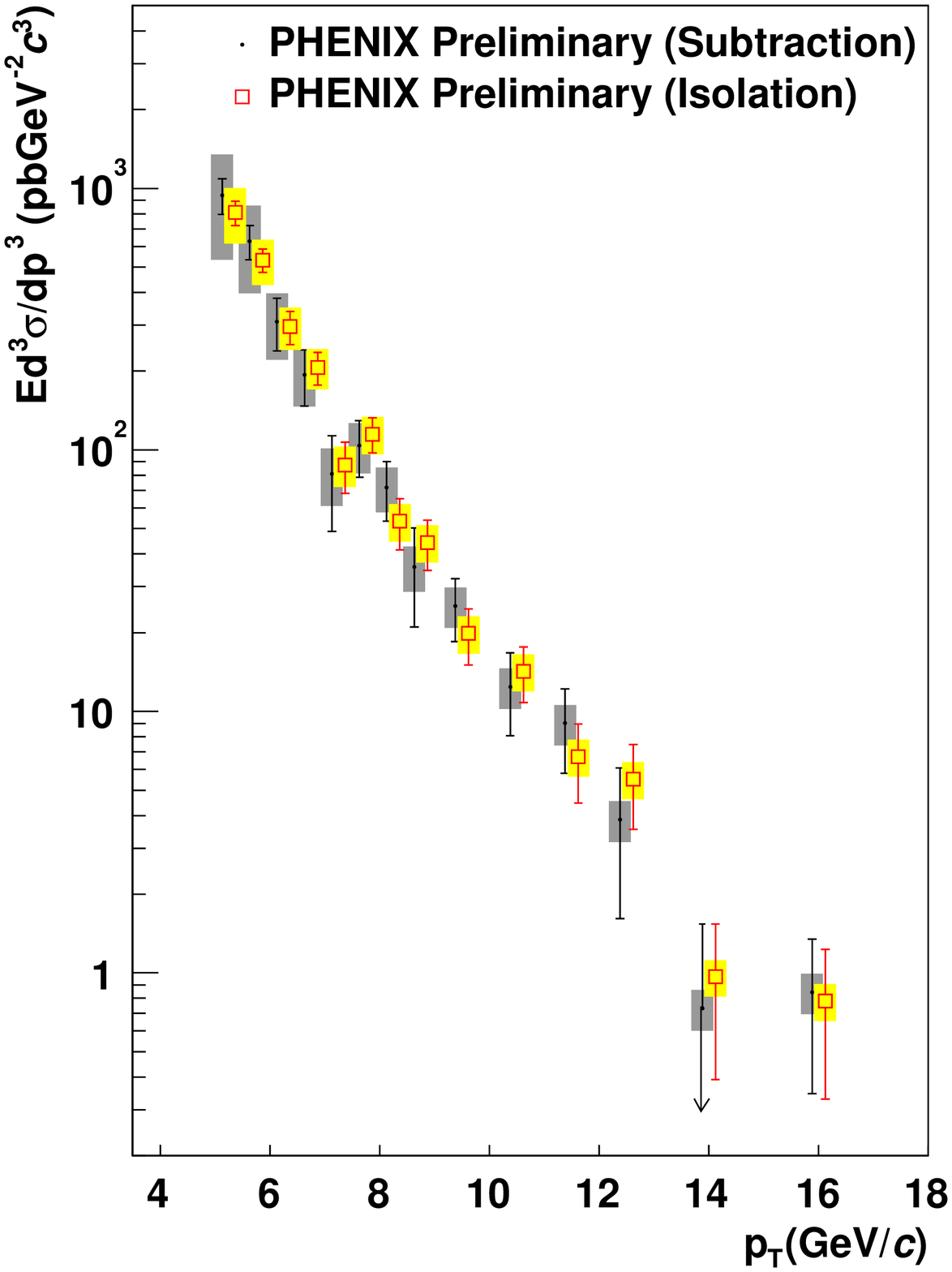}
\end{center}
\end{minipage} 
\caption{\label{fig:fig1_pp}PHENIX preliminary direct photon invariant
cross section \cite{Okada:2005in} in $p+p$ collisions at $\sqrt{s}$ =
200~GeV .  The left panel shows the result obtained with even-by-event
subtraction of decay photons. The curves represent a NLO pQCD
calculation for three different scales.  The right panel compares this
result to the result obtained by an isolation cut.}
\end{figure}

The first step in the direct photon analysis in Au+Au is to get a
clean inclusive photon sample.  To achieve this, non-photon
backgrounds, as from charged particles, have to be subtracted.  The
next step is to measure the $p_T$ spectrum of $\pi^0$'s and $\eta$'s.
From these spectra, the number of decay photons is determined in a
Monte Carlo calculation, which also considers contributions from other
hadronic decays.  Finally, the decay photon spectrum is subtracted
from the inclusive photon spectrum to obtain the direct photon
spectrum.  This method makes no attempt to identify direct photons on
an event-by-event basis.  Therefore it is called the subtraction
method or statistical method.  Further details can be found in
\cite{Adler:2005ig}.

In p+p collisions, where the multiplicity and the occupancy on the
detector are small, it is possible to tag and subtract decay photons
event by event.  In addition, an isolation cut, {\it i. e.} a cut on
the maximum energy allowed in a cone around a direct photon, allows
further suppression of decay photons on an event-by-event basis.  As
an additional feature the isolation cut helps to identify
fragmentation photons.  Further details on the $p+p$ analysis can be
found in \cite{Okada:2005in}.

For the measurement of direct photon production in $p+p$ collisions,
an integrated luminosity of 266 $\rm{nbarn}^{-1}$ sampled in the 2003
RHIC $p+p$ run was analyzed.  The Au+Au result is based on the 2002
run with an integrated luminosity of 24 $\rm{\mu barn}^{-1}$.
Photons were detected by the electromagnetic calorimeter (EMCal),
which is located in the two central arms of the PHENIX detector, each
covering $2*90^\circ$ in azimuth and a pseudorapidity ($\eta$) range
of $\pm0.35$.  For the $p+p$ measurement, only one arm was used.  The
energy calibration was checked by the position and width of the
$\pi^0$ invariant-mass peak.  The fraction of charged-particle
contamination was determined with the central-arm tracking detectors
which are located in front of the EMCal.

\begin{figure}[t]
\includegraphics[width=18pc]{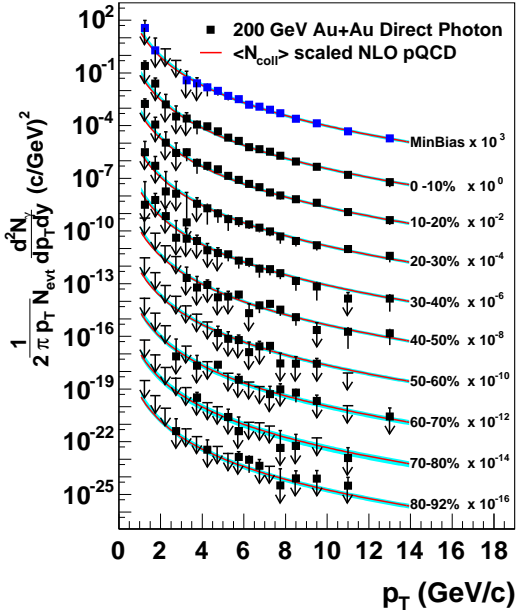}\hspace{2pc}%
\begin{minipage}[b]{18pc}\caption{\label{fig:fig3_spe}
Direct photon invariant yields \cite{Adler:2005ig} as a function of
$p_T$ for 9 centrality selections and minimum bias Au+Au collisions at
$\sqrt{s_{_{NN}}}$ = 200~GeV.  The error bars indicate the total
error.  Arrows indicate measurements consistent with zero yield with
the tail of the arrow indicating the 90\% confidence level upper
limit.  The solid curves are pQCD predictions.}
\end{minipage}
\end{figure}

\section{Direct Photon Cross Section in $p+p$ Collisions}

Figure \ref{fig:fig1_pp} shows the PHENIX preliminary direct photon
cross section measured in $\sqrt{s} = 200$\,GeV $p+p$ collisions as a
function of $p_T$ \cite{Okada:2005in}.  The left panel shows the
result obtained with event-by-event tagging and subtraction of decay
photons. The curves represent a NLO pQCD calculation for three
different scales \cite{Gordon:1993qc}.  The calculation agrees well
with the measurement.  The right panel of Fig. \ref{fig:fig1_pp}
compares this result to the result obtained by an isolation cut.  No
significant reduction of the direct photon yield by the isolation cut
is observed.  This suggests that either the contribution from
fragmentation photons is small or that the efficiency of the isolation
cut to discount fragmentation photons is low.  These preliminary
results substantially increase the statistical significance and the
$p_T$ reach of the published result from the 2002 run
\cite{Adler:2005qk}.

\begin{figure}[t]
\begin{center}
\includegraphics[width=0.5\linewidth]{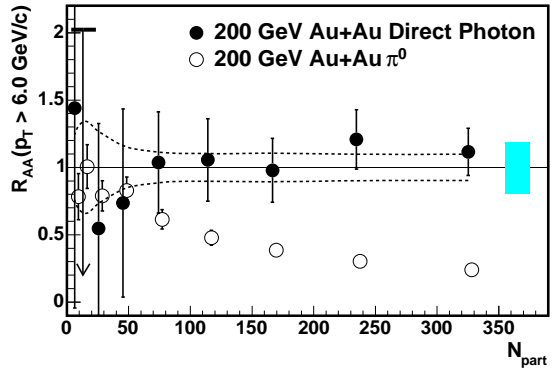}
\end{center}
\caption[]{$R_{AA}(p_T>6$GeV$/c)$ as a function of centrality
($N_{part}$) for direct photons (closed circles) and $\pi^0$'s (open
circles) \cite{Adler:2005ig}. The error bars indicate the total error
excluding the error on $\langle N_{coll} \rangle$ shown by the dashed
lines and the scale uncertainty of the NLO calculation shown by the
shaded region at the right.}
\label{fig:fig4_raa}
\end{figure}

\section{Direct Photon Yield in Au+Au Collisions}

The direct photon spectra measured in $\sqrt{s_{NN}} = 200$\,GeV Au+Au
collisions \cite{Adler:2005ig} are shown in Fig. \ref{fig:fig3_spe}
for different centrality selections.  They are compared to NLO pQCD
calculations \cite{Gordon:1993qc} scaled by the corresponding number
of binary nucleon collisions.  It can be seen that the pQCD
calculations provide a good description of the measured direct photon
spectra.

A common way to study possible medium effects in $AA$ collisions is
the \emph{nuclear modification factor} $R_{AA}$, {\it i. e.} the ratio
of the $AA$ invariant yields to the $NN$-collision-scaled $p+p$
invariant yields \cite{Adler:2003qi}.  The centrality dependence of
$R_{AA}$ for the integrated yield in $p_T>6$ GeV$/c$ for direct
photons (closed circles) is now compared to that of $\pi^0$'s (open
circles) in Fig.~\ref{fig:fig4_raa}.  The direct photon $p+p$ yield is
taken as the NLO pQCD calculation \cite{Gordon:1993qc} as in the
previous figure, while the $\pi^0$ $p+p$ yield is taken from the
measured $\pi^0$ yield~\cite{Adler:2003pb}.  The $R_{AA}$ trend
confirms the observation from above in more detail: the direct photon
production in Au+Au is consistent with the binary-scaled $p+p$ pQCD
calculation.  This is in sharp contrast \cite{Adler:2003qi} to the
centrality dependence of the $\pi^0$ $R_{AA}$, indicating that the
observed large suppression of high-$p_T$ hadron production in central
Au+Au collisions is dominantly a final-state effect due to parton
energy loss in the dense produced medium, rather than an initial-state
effect.

\section{Summary}

Direct photon production has been measured in $p+p$ and Au+Au
collisions at $\sqrt{s_{NN}}=200$ GeV.  In $p+p$, NLO pQCD
calculations agree well with the measurement.  The equally good
agreement between measurement and binary scaled pQCD calculations in
Au+Au suggests that nuclear modifications at mid-rapidity are
small. The result provides strong confirmation that the observed large
suppression of high $p_T$ hadron production in central Au+Au
collisions is dominantly a final-state effect.

\section*{References}
\bibliographystyle{unsrt.bst}

\bibliography{photons}

\end{document}